# ANALOG SIGNAL MULTIPLEXING SYSTEM FOR THE IOTA PROTON INJECTOR

Daniel R. MacLean[1,†], Dean R. Edstrom[1]
[1]Fermi National Accelerator Laboratory, Batavia, IL, United States

*Abstract*

The Fermilab Accelerator Science and Technology (FAST) Facility at FNAL is a dedicated research and development center focused on advancing particle accelerator technologies for future applications worldwide. Currently, a key objective of FAST Operations is to commission the 2.5 MeV IOTA Proton Injector (IPI) and enable proton injection into the Integrable Optics Test Accelerator (IOTA) storage ring. The low and medium-energy sections of the IPI include four frame-style dipole trims and two multi-function correctors with independently controlled coils, requiring readout of 32 analog channels for current and voltage monitoring in total. To reduce cost and optimize rack space within the PLC-based control system, a 32-to-4 analog signal multiplexing system was designed and implemented. This system enables real-time readback of excitation parameters from all magnetic correctors. This paper presents the design, construction, implementation, and performance of the multiplexing system.

## IPI OVERVIEW

The IOTA Proton Injector is located at the Fermi Accelerator Science and Technology Facility, a dedicated research and development facility focusing on the advancement of particle accelerator technology. The approximately ~12-Meter long IPI beamline will deliver 2.5 MeV protons to the IOTA storage ring for injection onto orbit, in order to carry out studies on nonlinear beam dynamics as well as to supplement ongoing research on electron beam cooling.

IPI is divided into four primary segments. The 50 kV Proton Source Stand produces beam within a Duoplasmatron Ion Source, which is then extracted into the Low-Energy Beam Transport (LEBT) section. Next, the proton beam is accelerated to the design energy of 2.5 MeV by a Radio-Frequency Quadrupole (RFQ) and is transported into a Medium-Energy Beam Transport (MEBT) section. The MEBT finally delivers the proton beam to IOTA where it is injected onto orbit for studies [1, 2].

## MAGNETIC ELEMENTS

Along the IPI beamline there are twelve focusing quadrupole magnets, two large bend dipoles, as well as several magnetic corrector elements comprised of four window-style horizontal-vertical (H-V) trim dipoles and two multi-function corrector (MFC) magnets with four independently powered coils each. The two bend dipoles define a 15° dogleg section for the purpose of improved matching into the IOTA ring. The H-V trims function as effectively two independent steering dipoles within a single package (two pairs of independently powered coils). MFCs can be configured to act as either a horizontal or vertical trim, as well as a skew-quadrupole element. Corrector elements, for the purposes of this system, refers to the trim dipoles and multi-function correctors. Table 1 summarizes all powered electromagnets used in IPI, and Figure 1 shows the corrector elements powered specifically by the system discussed here [2].

Table 1: IPI Beamline Magnetic Elements: Boldfaced Elements are Readout by the System Discussed Here

| Element Type | Magnet Count | # Powered Channels/Magnet |
| --- | --- | --- |
| Quadrupoles | 12 | 1 |
| Bend Dipoles | 2 | 1 |
| **Trim Dipoles** | **4** | **2** |
| **MFCs** | **2** | **4** |

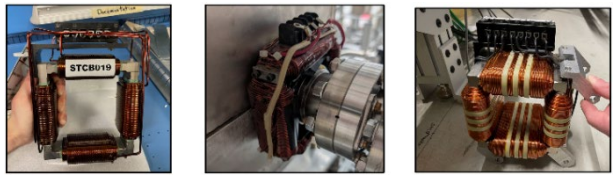

Figure 1: IPI magnet corrector elements, left to right: Dipole trim (Style #1), dipole trim (Style #2), multi-function corrector.

## POWER & READBACKS

Power is provided to the IPI magnetic corrector elements by a bipolar ±2 Amp current regulator chassis, which is in turn powered by two 33 Volt, 33 Amp DC bulk power supplies connected in-series and referenced to the system common at the center, providing two bipolar voltage rails about Ground of ±15 Vdc. 16 channels – i.e. individual magnet coils - are powered by this system in-total, each with corresponding current and voltage sense readbacks. Analog V/I Readbacks are output by the 2-Amp regulator chassis and routed back to the PLC-based magnet control system. These readbacks are DC levels scaled to the full range of voltage & current output of the regulator. The bipolar readback levels range from -10 to +10 Vdc and are derived from sense-resistors at the output terminals of each channel within the regulator chassis; the nominal scaling of the current readbacks for each individual channel is therefore 0.2 Amps/Volt.

The entire array of magnetic element power supplies for IPI, as well as the associated PLC-based control system, is housed within a single relay rack in the External Service Building (ESB) at the FAST Facility. With limited available analog-to-digital converter (ADC) channels at the IPI magnet control PLC, it was determined that multiplexing the V/I readbacks from the corrector element regulator chassis was necessary. This allowed the entire system to be readout with a single ADC card, vs. the eight that would be required with an un-multiplexed system, reducing overall cost and optimizing relay rack space.

## 32:4 MULTIPLEXING MODULE

Multiplexing the 32 analog readback signals for the IPI magnetic correction elements is accomplished using a circuit based primarily upon the 74HC4051 ('4051) 8-channel analog multiplexing/de-multiplexing integrated circuit (IC). The '4051 is produced by Texas Instruments and is readily available through all major electronics component distributers. Four 74HC4051 units are used in 'parallel' to produce a 4-channel multiplexed readback output, which is then passed into the ADC cards handling inputs to the Magnet Control PLC. See Figure 2 for a simplified overview of the entire system discussed in this section.

Before any signal processing takes place, the incoming V/I readbacks from the 2-Amp regulator chassis must be voltage-divided down to levels that are within the operating specifications of the 74HC4051. The '4051 analog inputs can only handle an input voltage range of -5 to 5 Vdc [3], thus the -10 Vdc to 10 Vdc analog readbacks from the regulator chassis are first passed through a fixed resistive voltage-divider stage, configured to drop the magnitude of the incoming unprocessed signals exactly by a factor of 2. Therefore, the full range of signals seen by the MUX stage of the circuit spans -5 to +5 Vdc, and the factor-of-two division must be accounted for in software, once the readbacks are fully calibrated for each channel within the PLC stage of the system

Power for the MUX module is provided by two onboard linear voltage regulators (LVRs), which establish +5 Vdc and -5 Vdc rails to control the '4051 ICs. The LVRs are of the LM78xx / LM79xx family, manufactured by Texas Instruments. Input power to the regulators is provided by the same ±15 Vdc rails from the bulk supplies used to power the Current Regulator Chassis for the Corrector Magnet system. A bi-polar power rail arrangement is required as the analog readbacks from the Regulator Chassis follow the polarity of the ±2 Amp output current to each magnet coil. Finally, a small DC cooling fan is mounted to the enclosure of the MUX module, in order to keep the two LVRs within their operational temperature range. The fan is powered directly from the input +15 Volt rail of the bulk power for the system.

The PLC used for the IPI Magnet System is manufactured by Automation Direct and is part of the Productivity 2000 series (which are used extensively at the FAST Facility across numerous systems). Input ADC modules for this system are P2-04AD 4-channel, 16-bit resolution models. The 3-bit Binary-Coded Decimal (BCD) control signals (S0,S1,S2) that switch/select the active input pin of each of the four '4051 ICs are generated by the Magnet Control PLC as well, using a P2-16TD-TTL Digital Output Module [4]. These control signals are distributed into the MUX module synchronously [3–5]. Each '4051 receives the same address-select control flag simultaneously, switching its active input pin synchronously with the other three MUX ICs [3]. These BCD control signals use a 5-Volt TTL level scheme and are switched at a nominal rate of 25 Hz – this represents the multiplexing rate of the readback system. This rate is also the upper-limit of the cycle-time setting for the Central Processing Unit (CPU) within the PLC that can be reliably implemented to readout the entire array of ADC channels into the Fermilab accelerator control system (ACNET) Modbus registers. The TTL control signals can, however, be manipulated via the ACNET control system for troubleshooting / testing purposes; the multiplexing rate can be adjusted, and the string of eight BCD control flags can be manually cycled to pass one readback channel at a time through each '4051 IC.

The four multiplexed output channels from the main signal processing stage of the circuit are passed through an operational amplifier (op-amp) based unity-gain buffer stage to ensure good isolation between the multiplexing circuit and the input ADC modules of the PLC system as the final stage of the signal processing chain. The Buffer module utilizes an LM324 quad op-amp produced by Texas Instruments and is unity-gain stable [6]. The MUX output buffer is installed directly adjacent to the multiplexing stage of the system within the IPI Magnet relay rack.

The initial prototype MUX module and buffer stage were constructed using general-purpose prototype soldered breadboards, housed within two separate enclosures, which were mounted to the Unistrut rails at the rear of the IPI magnet power / control rack. After the installation & commissioning of the original prototype module for IPI, a dedicated PCB-based version of the system was designed and constructed as well, currently serving as a backup system in the event of a failure with the original. The Rev. A board for the system includes both the multiplexing stage (and its associated power-handling section), as well as the unity-gain buffer stage on the same board, in contrast to the prototype system, where these two stages are on separate boards, within separate albeit physically close enclosures.

## FUTURE WORK

The same model ±2-Amp Current Regulator Chassis is used to power numerous magnetic elements within the IOTA Ring itself – including the sextupole and octupole correctors, and the same style multi-function correctors as in IPI. Given the effectiveness and relative simplicity / serviceability of the MUX Module implemented for the IOTA Proton Injector, there has been discussion of scaling the system to magnetic corrector element readback channels for the IOTA ring proper – especially in the case of future expansions and addition of elements for forthcoming experiments. The implementation would be essentially

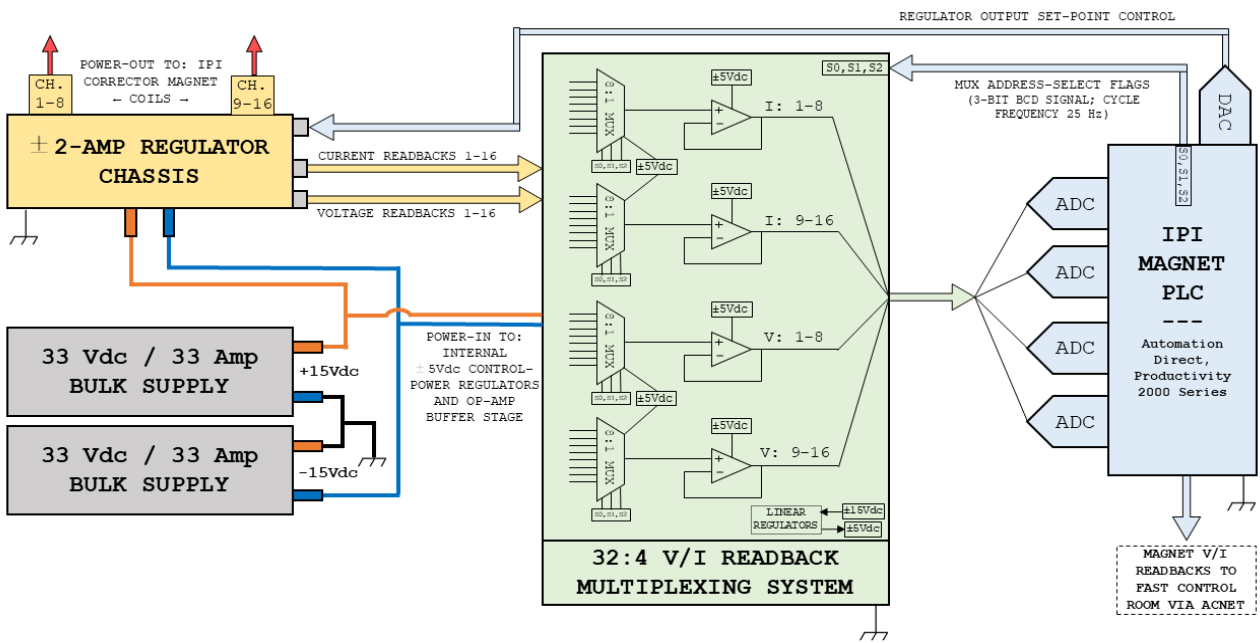

Figure 2: Simplified block-diagram of the entire power & readback system for the IPI magnetic corrector elements, including links between the bulk power supplies / Current Regulator Chassis as well as the PLC controlling the entire IPI magnet system.

identical, with dedicated PCBs and enclosures for each ±2-Amp Current Regulator Chassis.

## CONCLUSIONS

A 32:4 analog multiplexing module was designed, constructed, and implemented for the upcoming first Run of the IOTA Proton Injector, in order to more compactly and efficiently handle the relatively dense set of V/I readback signals from the set of magnetic correction elements. Initial tests have proven successful; the prototype MUX system has been used to fully power-test all correctors as well as to calibrate the readbacks originating from the corrector regulator chassis and properly interface them with the ACNET accelerator control system.

## ACKNOWLEDGEMENT

This manuscript has been authored by FermiForward Discovery Group, LLC under Contract No. 89243024CSC000002 with the U.S. Department of Energy, Office of Science, Office of High Energy Physics.

## DISCLAIMER

Neither the United States nor the United States Department of Energy, nor any of their employees, makes any warranty, express or implied, or assumes any legal liability or responsibility for the accuracy, completeness, or usefulness of any data, apparatus, product, or process disclosed, or represents that its use would not infringe privately owned rights.

## REFERENCES

[1] S. Antipov *et al.*, "IOTA (Integrable Optics Test Accelerator): facility and experimental beam physics program," Journal of Instrumentation, vol. 12, no. 03, pp. T03002–T03002, Mar. 2017. doi:10.1088/1748-0221/12/03/t03002

[2] D. Edstrom *et al.*, "IOTA Proton Injector Beamline Installation", in *Proc. IPAC'23*, Venice, Italy, May 2023, pp. 1737-1739. doi:10.18429/JACoW-IPAC2023-TUPA183

[3] *High speed CMOS logic analog multiplexers/demultiplexers,* Texas Instruments, datasheet 74HC4051, Nov. 1997; https://www.ti.com/lit/gpn/74HC4051

[4] *Digital Output Modules, P2-16TD-TTL,* Automation Direct, datasheet P2-16TD-TTL, Jun. 2024; https://cdn.automationdirect.com/static/specs/p216td-ttl.pdf

[5] *Analog Input Modules, P2-04AD,* Automation Direct, datasheet P2-04AD, Aug. 2024; https://cdn.automationdirect.com/static/specs/p204ad.pdf

[6] *LM324/LM2902B Quadruple Operational Amplifiers Datasheet,* Texas Instruments, datasheet SNOSBT3L, Jan. 2015, Rev. Nov 2021.